\documentclass[osajnl,twocolumn,showpacs,superscriptaddress,10pt]{revtex4-1} 
\usepackage{amsmath,amssymb,graphicx}

\begin{document}
\title{Surface Integral Method for the Second Harmonic Generation in Metal Nanoparticles}

\author{Carlo Forestiere}
\affiliation{ Department of Electrical Engineering and Information Technology, Universit\`{a} degli Studi di Napoli Federico II, via Claudio 21,
Napoli, 80125, Italy}
\affiliation{Department of Electrical and Computer Engineering \& Photonics Center, Boston University, 8 Saint Mary’s Street, Boston, Massachusetts}

\author{Antonio Capretti}
\affiliation{ Department of Electrical Engineering and Information Technology, Universit\`{a} degli Studi di Napoli Federico II, via Claudio 21,
Napoli, 80125, Italy}

\author{Giovanni Miano}\email{Corresponding author: miano@unina.it}
\affiliation{ Department of Electrical Engineering and Information Technology, Universit\`{a} degli Studi di Napoli Federico II, via Claudio 21,
Napoli, 80125, Italy}

\begin{abstract}
Second harmonic (SH) radiation in metal nanoparticles is generated  by both nonlocal-bulk and local-surface SH sources, induced by the
electromagnetic field at the fundamental frequency.  We propose a surface integral equation (SIE) method for evaluating the SH radiation generated by
metal nanoparticles with arbitrary shapes, considering all SH sources. We demonstrate
that the contribution of the nonlocal-bulk SH sources to the SH electromagnetic field can be  taken into account through equivalent
surface electric and magnetic
currents. We numerically solve the SIE problem by using the Galerkin method and the Rao-Wilton-Glisson basis functions in the framework of the
distribution theory.
The accuracy of the proposed method is verified by comparison with the SH-Mie analytical solution. As an example of a complex-shaped particle, we
investigate the SH scattering by a triangular nano-prism. This method paves the 
way for a better understanding of the SH generation process in arbitrarily shaped nanoparticles and can also have a high impact in the design of novel
nanoplasmonic devices with enhanced SH emission.
\end{abstract}

\maketitle 
\section{Introduction}

The study of nonlinear optical properties of noble metal nanoparticles (NPs) has become a very active research field, stimulated by both the maturity
of nanofabrication techniques and the appeal of the potential applications \cite{Kauranen12}. Noble metal nanoparticles support localized surface
plasmon (LSP)
resonances, consisting in the collective oscillations of the conduction electrons \cite{MaierBook}. LSPs can significantly enhance both the linear and
nonlinear scattering  \cite{Bouhelier2003}, enabling strong nonlinear optical effects at relatively low excitation powers. 

The second harmonic (SH) generation is a nonlinear optical process in which two photons of the same frequency, {\it i.e.} the fundamental frequency,
interact with the material and generate one photon of twice the frequency. SH generation from noble metal NPs takes place in the bulk of the particle
and on its surface. In particular, the electromagnetic field at the fundamental frequency induces nonlocal-bulk and local-surface SH sources, whereas
the local-bulk SH sources are absent due to the centrosymmetry of the material \cite{Heinz91}. The
local-surface SH sources are due to the symmetry breaking at the interface between the NP and the embedding medium \cite{Dadap99}. The relative
weights of the different nonlinear sources depend on the NP shape and
size, and on the optical properties of the metal at both the fundamental and the SH frequencies \cite{GuyotSionnest86}. 

Several analytical models have been developed to describe the SH scattering from centrosymmetric NPs of simple shapes such as spheres and
cylinders. In particular, Dadap et. al \cite{Dadap99} studied the SH radiation generated from the local-surface SH sources in a sphere with radius
much
smaller than the wavelength of the exciting light (Rayleigh limit).  This model was later extended in \cite{Mochan03,Dadap04}, considering also the
nonlocal-bulk sources. A full-wave theory of the SH generation by a sphere of centrosymmetric material featuring SH  local-surface sources has
been proposed in \cite{Pavlyukh04}. This approach has been recently extended to model aggregates of spheres \cite{Xu12} but, also in this case, only
the local-surface SH sources have been taken into account. A full-wave analytical theory has been also proposed for the SH scattering of
light by a cylinder of infinite length \cite{Valencia04}.

The last few decades have witnessed a continuous improvement of nanofabrication techniques and a large variety of complex shapes can
 be controllably achieved nowadays. This fact allows for an accurate tuning of the spectral position of the LSP resonance of a NP and stimulates the
demand for numerical tools to model the linear and non-linear behavior of
NPs with arbitrary
shapes. In fact, De Beer {\it et al.} \cite{deBeer11} presented a theoretical framework for solving the SH generation problem for particles of
arbitrary shape in the low contrast limit. Makitalo et. al. \cite{Makitalo11} introduced a surface integral equation (SIE) formulation for studying
the nonlinear scattering from arbitrarily shaped particles taking into account only the local-surface SH sources. Benedetti et. al.
\cite{Benedetti11} developed a volume integral equation (VIE) formulation based on the dyadic Green function for the study of non-linear
scattering properties of arbitrarily shaped particles due to both nonlocal-bulk and local-surface contributions. However, unlike SIE formulations,
the VIE formulations require the discretization of the entire NP
volume, resulting in an higher memory occupancy and longer computational time.

In this paper, we present a SIE formulation of the SH generation problem for three dimensional, arbitrarily shaped nanoparticles made of
centrosymmetric and lossy materials, including both the nonlocal-bulk and local-surface nonlinear polarization sources. The
nonlocal-bulk polarization sources play an important role and has to be considered in any realistic model of SH generation \cite{Wang09,
Benedetti11}.
 The problem is treated in the frequency domain under the undepleted-pump approximation.  In particular, we demonstrate that the contribution of the
nonlocal-bulk sources to the SH electromagnetic field can be taken into account by introducing fictitious electric and magnetic currents on
the particle boundary, leading to a great simplification of the numerical problem. The excitation vectors of the SIE system have been derived in the
framework of the distribution theory in order to correctly account for the discontinuity of the discrete
representation of the SH sources.

The paper is organized as follows. Section 2 presents the mathematical statement of the problem along with the expression of both the nonlocal-bulk
and the local-surface nonlinear polarization sources of the SH radiation. Then Section 3 is devoted to the SIE formulation and its numerical solution
through the Galerkin method. Section 4 deals with the validation of the method, accomplished through an extensive comparison with the full wave SH Mie
solution for nanoparticles with spherical shapes proposed in ref. \cite{CaprettiPRB}. Finally, the SH scattering from a gold triangular prism is
presented.

\section{Statement of the Problem}

In this section, we formulate the problem of SH generation from a lossy NP of arbitrary shape in a homogeneous
medium, when it is illuminated by a time-harmonic electromagnetic field at frequency $\omega$ . The domain of the
electromagnetic field is the entire
space $\mathbb{R}^3$, which is divided into the interior of the metal domain $\overset{\circ}{V}_i$, the external medium $\overset{\circ}{V}_e$ 
and the interface $S$.  The closed surface $S$ is oriented such that its normal ${\bf n}$ points outward. Furthermore, ${\bf r}$ denotes
the position vector with respect to an arbitrary reference point ${O}$. Since the intensities of the SH fields generated by noble metals NP are
always order of magnitude weaker than the intensities of the pump fields, we can assume that the SH fields do not couple back to the
fundamental fields ({\it undepleted-pump approximation}). Therefore, the non-linear electromagnetic problem can be decomposed in two linear scattering
problems at the fundamental frequency and at the SH frequency. The nonlinear response of the material is described by the SH
polarization sources induced by the fundamental electric field. 

We denote with $\left( {\bf E }_l^{\left( \omega \right)}, {\bf H }_l^{\left( \omega \right)} \right)$ the fundamental fields at frequency $\omega$ in
the region $\overset{\circ}{V}_l$ with  $l=i,e$,  and with $\left( {\bf E }_l^{\left( 2 \omega \right)}, {\bf H }_l ^{\left(2 \omega \right)} 
\right)$  the SH field at frequency $2\omega$. We use the convention ${\bf a} \left( {\bf r} \right) = \Re \left\{ {\bf A}^{\left( \Omega \right) }
\left( {\bf r} \right) \exp \left( j \Omega t \right) \right\} $ for representing a time harmonic electromagnetic field. Furthermore, we denote with 
$\left( {\bf E }_0^{\left( \omega \right)}, {\bf H }_0^{\left( \omega \right)} \right) $ the pump fields incident in $\overset{\circ}{V}_e$, with
$\left( \varepsilon_i \left\{ \Omega \right\}, \mu_i \right) $ the linear permittivity at frequency $\Omega$ and the permeability of the metal, and
with  and $\left( \varepsilon_e, \mu_e \right) $ the permeability and the permittivity of the embedding medium.
In order to calculate the SH radiation generated by the metal particle we have to evaluate: 1) the electric fields ${\bf E}_l^{\left( \omega
\right)}$  at the fundamental frequency induced by the pump electromagnetic field; 2) the bulk and surface nonlinear sources generated by ${\bf
E}_l^{\left( \omega \right)}$ and 3) the SH fields  $\left( {\bf E }_l^{\left(2 \omega \right)}, {\bf H }_l^{\left(2 \omega \right)} \right) $
generated by the nonlinear sources. 


Aiming at the solution of scattering problem at the fundamental frequency, we introduce the scattered fields $ \left(
{\bf{E}}_l^{\left(\omega, s \right)},{\bf{H}}_l^{\left(\omega, s \right)} \right)$ with $l=i,e$, defined as:
\begin{equation}
  \left\{
  \begin{aligned}
    {\bf{E}}_e^{\left(\omega, s \right)} &= {\bf{E}}_e^{ \left(\omega \right) }  - {\bf{E}}_0^{\left(\omega \right)} \\ 
    {\bf{H}}_e^{\left(\omega, s \right)} &= {\bf{H}}_e^{ \left(\omega \right) }  - {\bf{H}}_0^{\left(\omega \right)} 
  \end{aligned}
  \right.
  \, \mbox{in} \;  \overset{\circ}{V}_e   
  \quad
  \left\{
  \begin{aligned}
    {\bf{E}}_i^{\left(\omega, s \right)} &= {\bf{E}}_i^{ \left(\omega \right) }  \\ 
    {\bf{H}}_i^{\left(\omega, s \right)} &= {\bf{H}}_i^{ \left(\omega \right) } 
  \end{aligned}
  \right.
  \, \mbox{in} \; \overset{\circ}{V}_i   
 \notag
\end{equation}
The  scattered fields at the fundamental frequency can be determined by solving the problem:
\begin{equation}
   \label{eq:FundamentalProblem}
   \begin{aligned}
   &\left\{
   \begin{aligned}
     \nabla  \times {\bf{E}}_l^{\left(\omega, s \right)}  &=  - j\omega \mu _l {\bf{H}}_l^{\left(\omega, s \right)}  \\
     \nabla  \times {\bf{H}}_l^{\left(\omega, s \right)}  &= j\omega \varepsilon _l  \left\{ \omega \right\} {\bf{E}}_l^{\left(\omega, s \right)}  
   \end{aligned}
   \right.
   \quad \mbox{in} \;  \overset{\circ}{V}_l, \quad \mbox{with} \; l=i,e \; , \\
   &\left\{
   \begin{aligned}
      {\bf{n}} \times \left( {\bf{E}}_e^{\left(\omega, s \right)} - {\bf{E}}_i^{\left(\omega, s \right)} \right) &= -{\bf{n}}
\times {\bf{E}}_0^{\left(\omega \right)}  \\
      {\bf{n}} \times \left( {\bf{H}}_e^{\left(\omega, s \right)} - {\bf{H}}_i^{\left(\omega, s \right)} \right) &= -{\bf{n}} \times 
{\bf{H}}_0^{\left(\omega \right)}  
      \end{aligned}
   \right.
   \quad \mbox{on} \; S \; ,
 \end{aligned}
\end{equation}
with the radiation condition at infinity. Once the solution of problem \ref{eq:FundamentalProblem} has been obtained, the SH polarization sources can
be calculated. 

Noble metals are centrosymmetric materials, therefore the leading term of the bulk nonlinear polarization density ${\bf P}_b^{ \left( 2 \omega \right)
} $ is:
\begin{equation}
   {\bf P}_b^{ \left( 2 \omega \right) } = \varepsilon_0 \overset{\leftrightarrow}{\chi}_b^{\left( 2 \right)}: {\bf E}_i^{\left( \omega \right)}
\left( \nabla {\bf
E}_i^{ \left( \omega \right) } \right) \quad \text{for} \; {\bf r} \in \overset{\circ}{V}_i
   \label{eq:BulkPolarization0}
\end{equation}
where $ \overset{\leftrightarrow}{\chi}_b^{\left( 2 \right)}$  is the quadrupolar contribution to the second-order nonlinear bulk susceptibility of
the metal. In the case of isotropic and  homogeneous material the relation \ref{eq:BulkPolarization0} becomes:
\begin{equation}
   {\bf P}_b^{ \left( 2 \omega \right) } = \varepsilon_0 \gamma  \nabla \left( {\bf E}_i^{\left( \omega \right)} \cdot  {\bf
E}_i^{ \left( \omega \right) } \right) + \varepsilon_0 \delta' \left( {\bf E}_i^{\left( \omega \right)} \cdot  \nabla   \right)
{\bf E}_i^{
\left(
\omega \right) },
   \label{eq:BulkPolarization1}
\end{equation}
where $\gamma$ and $\delta'$ are material parameters \cite{Bloembergen68}. 
By using the vectorial identity $ \frac{1}{2} \nabla \left( {\bf A} \cdot {\bf A} \right) = {\bf A} \times \left( \nabla \times {\bf A} \right) +
\left( {\bf
A} \cdot \nabla \right) {\bf A}$ together with the Maxwell equations, we have:
\begin{equation}
   {\bf P}_b^{ \left( 2 \omega \right) }={\bf P}_{\gamma'}^{ \left( 2 \omega \right)} + {\bf P}_{\delta'}^{ \left( 2 \omega \right) }
   \label{eq:BulkPolarization2}
\end{equation}
where ${\bf P}_{\gamma'}^{ \left( 2 \omega \right)} = \varepsilon_0 \gamma'  \nabla \left( {\bf E}_i^{\left( \omega \right)} \cdot  {\bf
E}_i^{ \left( \omega \right) } \right) $, $ {\bf P}_{\delta'}^{ \left( 2 \omega \right) } = \delta' j \omega /c^2   {\bf
E}_i^{\left( \omega \right)} \times  {\bf H}_i^{\left( \omega \right)} $, and
$\gamma' = \gamma + \frac{\delta'}{2}$. Although eqs. \ref{eq:BulkPolarization1} and  \ref{eq:BulkPolarization2} are mathematically
equivalent, the latter is easier to treat numerically.

The local-surface SH polarization sources ${\bf P}_S^{ \left( 2 \omega \right) }$ are:
\begin{equation}
   {\bf P}_S^{ \left( 2 \omega \right) } = \varepsilon_0 \overset{\leftrightarrow}{\chi}_S^{\left( 2 \right)}: {\bf E}_i^{\left( \omega \right)}{\bf
E}_i^{\left(
\omega
\right)},
   \label{eq:SurfacePolarization0}
\end{equation}
where $\overset{\leftrightarrow}{\chi}_S^{\left( 2 \right)}$ is the second-order surface nonlinear susceptibility. Since the NP surface
has an isotropic symmetry with a mirror plane perpendicular to it, the surface-susceptibility tensor $\overset{\leftrightarrow}{\chi}_S^{\left( 2
\right)}$ has only three non-vanishing and independent elements, ${\chi}_{\perp \perp \perp}^{\left( 2 \right)}$,
${\chi}_{\perp \parallel \parallel}^{\left( 2 \right)}$  and ${\chi}_{\parallel \perp \parallel}^{\left( 2 \right)} = {\chi}_{\parallel \parallel
\perp}^{\left( 2 \right)}$, where $\perp$ and $\parallel$ refer
to the orthogonal and tangential components to the nanoparticle surface. Therefore, we have:
\begin{multline}
   {\bf P}_S^{ \left( 2 \omega \right) } = \epsilon_0 \left[  {\chi}_{\perp \perp \perp}^{\left( 2 \right)} {\bf n} {\bf n} {\bf n} +
{\chi}_{ \perp \parallel  \parallel}^{\left( 2 \right)} \left({\bf n}  {\bf t}_1 {\bf t}_1 + {\bf n} {\bf t}_2  {\bf
t}_2 \right)\right. + \\
\left. {\chi}_{ \parallel \perp \parallel}^{\left( 2 \right)} \left( {\bf t}_1 {\bf n} {\bf t}_1 + {\bf t}_2 {\bf n}
{\bf t}_2 \right) \right]: {\bf E}_i^{\left( \omega \right)}{\bf E}_i^{\left( \omega \right)},
   \label{eq:SurfacePolarization1}
\end{multline}
where $\left( {\bf t}_1, {\bf t}_2, {\bf n} \right)$  is a system of three orthonormal vectors locally defined on the
particle surface.  The contribution of tangential and normal component of the surface nonlinear polarization are taken into account by introducing the
surface electric ${\boldsymbol\pi}_{S}^{\left( e \right) }$ and magnetic ${\boldsymbol\pi}_{S}^{\left( m \right)}$ current densities
\cite{Heinz91}:
\begin{subequations}
   \begin{align}
   \label{eq:SurfaceElectricCurrent_SH} 
   {\boldsymbol\pi}_{S}^{\left( e \right) } &= j 2 \omega  {\bf P}_{t}^{S}, \\
   \label{eq:SurfaceMagneticCurrent_SH}
  {\boldsymbol\pi}_{S}^{\left( m \right) } &=  \frac{1}{\varepsilon_0} {\bf n} \times \nabla_S { P}_{n}^{S},
  \end{align}
\end{subequations}
where ${\bf P}_{t}^{S}  = - {\bf n} \times {\bf n} \times {\bf P}_S^{\left( 2\omega \right)}$ and  $  { P}_{ n}^{S} = {\bf P}_S^{\left( 2\omega
\right)} \cdot {\bf n}.  $

We determine the SH fields by solving non-homogeneous problem:
\begin{subequations}
\begin{align}
   \label{eq:SHG_Maxwell_e}
&  \left\{
  \begin{aligned}
     \nabla  \times {\bf {E} }_e^{ \left( 2\omega \right) }  &=  - j   2\omega  \mu _e {\bf {H} }_e^{ \left( 2\omega \right)
} \\
     \nabla  \times {\bf {H}}_e^{ \left( 2\omega \right) }  &=    j  2\omega \varepsilon _e {\bf {E} }_e^{ \left( 2\omega
\right) }
  \end{aligned}
  \right.
  \; \mbox{in} \;  \overset{\circ}{V}_e ,
  \\
\label{eq:SHG_Maxwell_i}
&
   \left\{
  \begin{aligned}
     \nabla  \times {\bf {E} }_i^{ \left( 2\omega \right) }  &=  - j 2 \omega \mu _i \, {\bf {H}}_i^{ \left( 2\omega
\right)}   \\
     \nabla  \times {\bf {H} }_i^{ \left( 2\omega \right)}  &=    j 2 \omega \varepsilon _i \left\{ 2\omega \right\} {\bf
{E} }_i^{ \left(
2\omega \right)} +  j 2 \omega  \left( {\bf P}_{\gamma'} + {\bf P}_{\delta'} \right)
  \end{aligned}
  \right.
  \; \mbox{in} \;  \overset{\circ}{V}_i , \\
 \label{eq:SHG_BoundaryCondition}
 &  \left\{
   \begin{aligned}
     {\bf n} \times \left( { {\bf {E} }_e^{ \left( 2\omega \right) }  - {\bf {E}}_i^{ \left( 2\omega \right) } } \right) &=  -
{\boldsymbol\pi}_{S}^{\left( m \right) }  \\
     {\bf n} \times \left( { {\bf {H} }_e^{ \left( 2\omega \right) }  - {\bf {H}}_i^{ \left( 2\omega \right) } } \right) &= 
{\boldsymbol\pi}_{S}^{\left( e \right) } 
   \end{aligned}
   \right.   \;\mbox{on} \; S \; ,
   \end{align}
\end{subequations}
with the radiation condition at infinity.
Due to the linearity of the problem \ref{eq:SHG_Maxwell_e}-\ref{eq:SHG_BoundaryCondition}, its general solution is the sum of the solution of the
homogeneous equation and one particular solution of the inhomogeneous equation. The eq. \ref{eq:SHG_Maxwell_i} is satisfied by the
particular solution:
\begin{equation}
\left\{
\begin{aligned}
   {\bf E}_{part} &= {\bf {E}}_{\gamma'}^{ \left( 2\omega \right) } + {\bf {E}}_{\delta'}^{ \left( 2\omega \right) }  \\
   {\bf H}_{part} &= {\bf {H}}_{\delta'}^{ \left( 2\omega \right) },
\end{aligned}
\right.
\end{equation}
where:
\begin{equation}
{\bf {E}}_{\gamma'}^{ \left( 2\omega \right) } =
        -  \frac{\varepsilon_0 \gamma'}{\varepsilon_i\left\{ 2\omega \right\} }  \nabla \left(
{\bf E}_i^{ \left( \omega\right) } \cdot {\bf E}_i^{ \left( \omega\right) } \right),
\end{equation}
\begin{multline}
{\bf {E}}_{\delta'}^{ \left( 2\omega \right) } = - j \omega \mu_0 \iiint_{V_i} {\bf J}_{\delta'}  \left( {\bf r}' \right) g_i^{\left( 2 \omega\right)}
 \left( {\bf r} - {\bf r}' \right) dV'  \\
        - \frac{1}{\varepsilon_i \left\{ 2\omega \right\}} \nabla  \left[ \iiint_{V_i} \rho_{\delta'} \left({\bf r}' \right)   g_i^{\left( 2
\omega\right)} \left( {\bf r} - {\bf r}' \right) dV' \right. \\ 
 + \left. \iint_S \eta_{\delta'} \left({\bf r}' \right)  g_i^{\left( 2 \omega\right)}  \left( {\bf r} - {\bf r}' \right) dS' \right] ,
\label{eq:E_Delta1}
\end{multline}
\begin{equation}
{\bf {H}}_{\delta'}^{ \left( 2\omega \right) }  = \nabla \times \iiint_{V_i} {\bf J}_{\delta'}  \left({\bf r}'
\right) g_i^{\left( 2 \omega\right)} \left( {\bf r} - {\bf r}' \right) dV' .
\label{eq:H_Delta1}
\end{equation}
The volumetric current density ${\bf J}_{\delta'}$, the volumetric charge density $\rho_{\delta'} $ and
the surface charge density $\eta_{\delta'} $ are given by:
\begin{equation}
\begin{aligned}
    {\bf J}_{\delta'}  \left({\bf r}' \right)  &= j 2 \omega  {\bf P}_{\delta'}, \\
        \rho_{\delta'}  \left({\bf r}' \right)  &= - \nabla \cdot  {\bf P}_{\delta'}, \\
        \eta_{\delta'}  \left({\bf r}' \right)  &=    {\bf n} \cdot  {\bf P}_{\delta'},
\end{aligned}
\end{equation}
where the quantity ${\bf P}_{\delta'}$ has been defined in eq.  \ref{eq:BulkPolarization2} and $g_i^{\left( \Omega\right)}$ is the homogeneous space
Green's function in the internal medium, {\it i.e.} $   g_i^{\left( 2 \omega\right)} \left( {\bf r} - {\bf r}' \right) = \frac{ e^{- j k_i \left\{
2\omega \right\} \left|{\bf r} - {\bf r}' \right|} }{4 \pi \left|{\bf r} - {\bf r}' \right| } $, $k_i \left\{ 2\omega \right\}=2\omega \sqrt{\mu_l
\varepsilon_i \left\{ 2\omega \right\} }$.
Therefore, the solution of the non-homogeneous problem defined by eqs. \ref{eq:SHG_Maxwell_e}-\ref{eq:SHG_BoundaryCondition} can be written as:
\begin{equation}
   \left\{
   \begin{aligned}
      {\bf {E}}_i^{ \left( 2\omega \right) }  &= {\bf \tilde{E}}_i^{ \left( 2\omega \right) } +{\bf {E}}_{\gamma'}^{ \left( 2\omega \right) } +{\bf
{E}}_{\delta'}^{ \left( 2\omega \right) }  \\
      {\bf {H}}_i^{ \left( 2\omega \right) }  &= {\bf \tilde{H}}_i^{ \left( 2\omega \right) } +{\bf
{H}}_{\delta'}^{ \left( 2\omega \right) }
   \end{aligned}
   \right.
   \qquad
   \left\{
   \begin{aligned}
      {\bf {E}}_e^{ \left( 2\omega \right) }  &= {\bf \tilde{E}}_e^{ \left( 2\omega \right) } \\
      {\bf {H}}_e^{ \left( 2\omega \right) }  &= {\bf \tilde{H}}_e^{ \left( 2\omega \right) }
   \end{aligned}
   \right.
\end{equation}
where  $ \left( {\bf \tilde{E}}_l^{ \left( 2\omega \right) }, {\bf \tilde{H}}_l^{ \left( 2\omega \right) } \right) $ with $l=i,e$ is the
solution of the problem:
\begin{equation}
  \left\{
  \begin{aligned}
     \nabla  \times {\bf \tilde{E} }_l^{ \left( 2\omega \right) }  &=  - j  2\omega \mu _l \, {\bf \tilde{H}}_l^{ \left( 2\omega
\right)}   \\
     \nabla  \times {\bf \tilde{H} }_l^{ \left( 2\omega \right)}  &=    j  2\omega \varepsilon _l \left\{ 2\omega \right\} {\bf
\tilde{E} }_l^{ \left( 2 \omega \right)} 
  \end{aligned}
  \right.
  \quad \mbox{in} \;  \overset{\circ}{V}_l  \quad \mbox{with} \; l=i,e \; ,   
\end{equation}
\begin{equation}
   \left\{
   \begin{aligned}
     {\bf n} \times \left( { {\bf \tilde{E} }_e^{ \left( 2\omega \right) }  - {\bf \tilde{E}}_i^{ \left( 2\omega \right) } } \right) &=  -
{\boldsymbol\pi}_{S}^{ \left( m \right) }- {\boldsymbol\pi}_{\gamma'}^{ \left( m \right) } - {\boldsymbol\pi}_{\delta'}^{ \left(
m\right) }  \\
     {\bf n} \times \left( { {\bf \tilde{H} }_e^{ \left( 2\omega \right) }  - {\bf \tilde{H}}_i^{ \left( 2\omega \right) } } \right) &= 
{\boldsymbol\pi}_{S}^{ \left( e \right) } + {\boldsymbol\pi}_{\delta'}^{ \left( e \right) } 
   \end{aligned}
   \right.
  \; \mbox{on} \; S \; ,
   \label{eq:SHG_Problem}
\end{equation}
with the radiation condition at infinity, where:
\begin{equation}
    \begin{aligned}    
    {\boldsymbol\pi}_{\gamma'}^{\left( m \right)} &=  \frac{1}{\varepsilon_0} {\bf n} \times
\nabla_S P_{n}^{ \gamma' }, \\
    {\boldsymbol\pi}_{\delta'}^{\left( m \right)} &= -{\bf n} \times {\bf {E}}_{\delta'}^{ \left( 2\omega \right) } \\
    {\boldsymbol\pi}_{\delta'}^{\left( e \right)} &=  {\bf n} \times {\bf {H}}_{\delta'}^{ \left( 2\omega \right) }
\end{aligned}
    \label{eq:BulkMagneticCurrent_SHG}
\end{equation}
and 
\begin{equation}
 { P}_{n}^{ \gamma'}  =  \frac {\varepsilon_0^2 \gamma'}{\varepsilon_i \left\{ 2\omega
\right\}  } \left(
{\bf E}_i^{ \left( \omega\right) } \cdot {\bf E}_i^{ \left( \omega\right) } \right).
\label{eq:BulkMagneticPolarization_SHG}
\end{equation}
In conclusion,  both nonlocal-bulk sources can be equivalently treated by using superficial electric and magnetic currents. In particular, the source
arising from $\gamma'$ can be associated to a fictitious surface polarization ${ P}_{n}^{ \gamma' } {\bf n}$, as already shown in \cite{Sipe87}.
Unlike ref. \cite{Sipe87}, this approach has also the advantage of the determination of the electromagnetic field due to the
nonlocal-bulk sources  ${{\bf P}}_{ \gamma' }^{\left( 2 \omega \right)}$ both outside and inside the particle. 

\section{Surface Integral Equations}
Let us define, for the same geometry described in the previous section, an auxiliary scattering problem that includes both the
fundamental and the SH problems as special cases. Let ${\boldsymbol\pi}_0^{\left( \Omega, e \right)}$, ${\boldsymbol\pi}_0^{\left( \Omega, m \right)}$
be the impressed electric and magnetic sources on $S$. We denote with $\left( {\bf{E}}_l^{\left( \Omega, s \right)},{\bf{H}}_l^{\left(\Omega,
s\right)}
\right)$ $l=i,e$ the fields solution of the problem:
\begin{equation}
\begin{aligned}
   \left\{
   \begin{aligned}
     \nabla  \times {\bf{E}}_l^{\left(\Omega, s \right)}  &=  - j\Omega \mu _l {\bf{H}}_l^{\left(\Omega, s \right)}  \\
     \nabla  \times {\bf{H}}_l^{\left(\Omega, s \right)}  &= j\Omega \varepsilon _l \left\{ \Omega \right\} {\bf{E}}_l^{\left(\Omega, s \right)}  
   \end{aligned}
   \right.
   \quad \mbox{in} \;  \overset{\circ}{V}_l \quad \mbox{with} \; l=i,e \; , 
    \\
   \left\{
   \begin{aligned}
      {\bf{n}} \times \left( {\bf{E}}_e^{\left(\Omega, s \right)} - {\bf{E}}_i^{\left(\Omega, s \right)} \right) &= -
{\boldsymbol\pi}_0^{\left(\Omega, m
\right) } \\
      {\bf{n}} \times \left( {\bf{H}}_e^{\left(\Omega, s \right)} - {\bf{H}}_i^{\left(\Omega, s \right)} \right) &=  +
{\boldsymbol\pi}_0^{\left(\Omega, e
\right) }
      \end{aligned}
   \right.
   \quad \mbox{on} \; S \; ,
   \label{eq:AuxiliaryProblem}
\end{aligned}
\end{equation}
with the radiation condition at infinity. We obtain the fundamental problem described of eq. \ref{eq:FundamentalProblem},
by substituting $\Omega = \omega$ and
\begin{equation}
  \left\{
  \begin{aligned}
      {\boldsymbol\pi}_0^{\left(\omega, e \right) } & = -{\bf{n}} \times  {\bf{H}}_0^{\left(\omega \right)},  \\
      {\boldsymbol\pi}_0^{\left(\omega, m \right) } & = {\bf{n}} \times  {\bf{E}}_0^{\left(\omega \right)}. 
  \end{aligned}
  \right.
  \label{eq:SetCurrentFund}
\end{equation}
Analogously, we can obtain the SH problem described by eq. \ref{eq:SHG_Problem} by substituting $\Omega = 2 \omega$ and
\begin{equation}
  \left\{
  \begin{aligned}
      {\boldsymbol\pi}_0^{\left( 2 \omega , e \right) } & =  {\boldsymbol\pi}_{S}^{\left( e \right) }+{\boldsymbol\pi}_{\delta'}^{\left( e \right) }, 
\\
      {\boldsymbol\pi}_0^{\left( 2 \omega , m \right) } & = {\boldsymbol\pi}_{S}^{\left( m \right) } + {\boldsymbol\pi}_{\delta'}^{ \left( m \right)
}+ {\boldsymbol\pi}_{\gamma'}^{\left( m \right)}.
  \end{aligned}
  \right.
  \label{eq:SetCurrentSHG}
\end{equation}

We now derive the surface integral formulation for the problem of eq. \ref{eq:AuxiliaryProblem}. By invoking the Love's equivalent principle
\cite{HarringtonBookMoM} for the exterior medium, the region $\overset{\circ}V_i$ is filled up with the
same material of the region $\overset{\circ}V_e$, and the sources are removed. The equivalent currents $ \left( {\bf j}_e^{\left( \Omega, e \right)},
{\bf j}_e^{\left(\Omega, m \right)} \right)$, positioned on the external page $S_e$ of the surface $S$, produce the field $\left( {\bf E}_e^{
\left(\Omega,s \right) }, {\bf H}_e^{\left(\Omega,s \right)} \right)$ in the region $\overset{\circ}{V}_e$, and null fields in the region
$\overset{\circ}{V}_i$, {\it i.e.}:
\begin{equation}
\begin{aligned}
   \mathcal{E}_e^{\left( \Omega \right) } \left\{ {\bf j}_e^{\left(\Omega, e \right)}, {\bf j}_e^{\left(\Omega, m \right)} \right\}
\left( {\bf{r}} \right) =
 \begin{cases}
     {\bf{0}}  & \text{if $ {\bf r} \in \overset{\circ}{V}_i$,} \\
     {\bf E}_e^{\left( \Omega,s \right) } \left( {{\bf{r}} } \right) & \text{if ${\bf r} \in \overset{\circ}{V}_e$,} \\
\end{cases} \\
\mathcal{H}_e^{\left( \Omega \right) } \left\{ {{\bf j}_e^{\left(\Omega, e \right)} ,{\bf j}_e^{\left(\Omega, m \right)} }
\right\}\left( {{\bf{r}} } \right)  =
 \begin{cases}
    {\bf{0}} & \text{if $ {\bf r} \in {\Omega}_i$,} \\
   {\bf H}_e^{\left( \Omega,s \right) } \left( {{\bf{r}} } \right) & \text{if ${\bf r} \in \overset{\circ}{V}_e$,} \\
\end{cases} 
\end{aligned}
   \label{eq:LoveExterior}
\end{equation}
where the sources $ \left( {\bf j}_e^{\left(\Omega, e \right)}, {\bf j}_e^{\left(\Omega, m \right)} \right)$ are given by:
\begin{equation}
  \left\{
  \begin{aligned}
     {\bf j}_e^{\left(\Omega, e \right)}  &=  +  \left. {\bf{n}} \times  {\bf H}_e^{\left( \Omega,s \right) } \right|_{S_e},  \\ 
     {\bf j}_e^{\left(\Omega, m \right)}  &=  - \left. {\bf{n}} \times  {\bf E}_e^{\left( \Omega,s \right) } \right|_{S_e}. 
  \end{aligned}
  \right.
  \label{eq:Je_Def}
\end{equation}
An equivalent problem may be set up also for the region  $\overset{\circ}{V}_i$. In this case the region $\overset{\circ}{V}_i$ is filled up with the
material with electromagnetic parameters $ \left( \varepsilon_i \left\{ \Omega \right\}, \mu_i \right) $. The equivalent
currents $ \left( {\bf
j}_i^{\left(\Omega, e \right)}, {\bf j}_i^{\left(\Omega, m \right)} \right)$, defined on the internal side $S_i$ of the surface $S$, produce the
original fields in the region $\overset{\circ}{V}_i$ and null field in the region $\overset{\circ}{V}_e$, {\it i.e.} 
\begin{equation}
\begin{aligned}
  \mathcal{E}_i^{\left(\Omega \right)} \left\{ {\bf j}_i^{\left(\Omega, e \right)}, {\bf j}_i^{\left(\Omega, m \right)} \right\}
\left( {{\bf{r}} } \right)  =&
 \begin{cases}
    {\bf E}_i^{\left(\Omega,s\right)} \left( {{\bf{r}} } \right)  & \text{if $ {\bf r} \in {\overset{\circ}{V}}_i$} \\
   {\bf{0}}                                            & \text{if ${\bf r} \in {\overset{\circ}{V}}_e$} \\
\end{cases}, \\ 
  \mathcal{H}_i^{\left(\Omega \right)} \left\{ {\bf j}_i^{\left(\Omega, e \right)},{\bf j}_i^{\left(\Omega, m \right)} \right\}
\left( {{\bf{r}} } \right)  = &
 \begin{cases}
    {\bf H}_i^{\left( \Omega,s \right) } \left( {{\bf{r}} } \right)  & \text{if $ {\bf r} \in {\overset{\circ}{V}}_i$} \\
    {\bf{0}}                                         & \text{if ${\bf r} \in {\overset{\circ}{V}}_e$} \\
\end{cases},
\end{aligned}
   \label{eq:LoveInterior}
\end{equation}
where the sources $ \left( {\bf j}_i^{\left(\Omega, e \right)}, {\bf j}_i^{\left(\Omega, m \right)} \right)$ are given by:
\begin{equation}
  \left\{
  \begin{aligned}
     {\bf j}_i^{\left(\Omega, e \right)}  &=  - \left. {\bf{n}} \times  {\bf H}_i^{\left(\Omega,s\right)} \right|_{S_i}  \\ 
     {\bf j}_i^{\left(\Omega, m \right)}  &=  + \left. {\bf{n}} \times  {\bf E}_i^{\left(\Omega,s\right)} \right|_{S_i} 
  \end{aligned} .
  \right.
  \label{eq:Ji_Def}
\end{equation}
Furthermore, the sets of equivalent currents  $ \left( {\bf j}_e^{\left(\Omega, e \right)}, {\bf j}_e^{\left(\Omega, m \right)} \right)$  and  $
\left( {\bf j}_i^{\left(\Omega, e \right)}, {\bf j}_i^{\left(\Omega, m \right)} \right)$ are not independent. In fact, combining the boundary
conditions of the problem \ref{eq:AuxiliaryProblem}  and the definitions \ref{eq:Je_Def} and \ref{eq:Ji_Def} we obtain:
\begin{equation}
   \left\{
  \begin{aligned}
     {\bf j}_i^{\left(\Omega, e \right)} + {\bf j}_e^{\left(\Omega, e \right)} &=  {\boldsymbol \pi}_0^{\left(\Omega, e \right)}, \\
     {\bf j}_i^{\left(\Omega, m \right)} + {\bf j}_e^{\left(\Omega, m \right)} &= {\boldsymbol \pi}_0^{\left(\Omega, m \right)}.
  \end{aligned}
\right.
\label{eq:IntExt}
\end{equation}

We now consider the limit of eqs. \ref{eq:LoveExterior} and \ref{eq:LoveInterior} as the point ${\bf r}$ approaches the surface
$S$ from its external face $S_e$ and internal face $S_i$. Projecting the resulting equations on the surface $S$ we obtain:
\begin{subequations}
  \begin{align}
    \label{eq:T-EFIEe}
    \mathcal{E}_e^{\left(\Omega, t \right)} \left\{ 
    {\bf j}_e^{\left( \Omega, e \right)} , {\bf j}_e^{\left(\Omega, m \right)} \right\}    - {\bf{n}} \times {\bf j}_e^{\left(
\Omega, m \right)} &= {\bf 0}, \\
    \label{eq:T-MFIEe}
  \mathcal{H}_e^{\left( \Omega, t \right)} 
   \left\{ {\bf j}_e^{\left(\Omega, e \right)}, {\bf j}_e^{\left(\Omega, m \right)} \right\} +   {\bf n} \times {\bf
j}_e^{\left(\Omega, e \right)} &= {\bf 0}, \\
    \label{eq:T-EFIEi}
 \mathcal{E}_i^{\left(\Omega,t \right)} \left\{    {\bf j}_i^{ \left( \Omega, e  \right) }, {\bf j}_i^{\left(
\Omega, m \right)} \right\}     + {\bf{n}} \times {\bf j}_i^{\left(
\Omega, m \right) } &= {\bf 0}, \\
    \label{eq:T-MFIEi}
   \mathcal{H}_i^{\left(\Omega, t \right) } 
    \left\{ {\bf j}_i^{\left(\Omega, e \right)} , {\bf j}_i^{\left(\Omega, m \right)} \right\}  - {\bf n} \times {\bf j}_i^{\left(\Omega, e \right)} 
&= {\bf 0}, 
  \end{align}
\end{subequations}
where we have defined the operators:
\begin{equation}
 \begin{aligned}
  \mathcal{E}_l^{\left(\Omega, t \right)} \left\{ {\bf j}^{\left( e \right)}, {\bf j}^{\left( m \right)}  \right\} &= \left. - {\bf n} \times
{\bf n}
\times \mathcal{E}_l^{\left(\Omega \right)} \left\{ {\bf j}^{\left( e \right)},{\bf j}^{\left( m \right)}  \right\} \right|_{S_l}, \\
  \mathcal{H}_l^{\left(\Omega, t \right)} \left\{ {\bf j}^{\left( e \right)}, {\bf j}^{\left( m \right)}  \right\} &= \left. - {\bf n} \times
{\bf n}
\times \mathcal{H}_l^{\left(\Omega \right)} \left\{ {\bf j}^{\left( e \right)},{\bf j}^{\left( m \right)}  \right\} \right|_{S_l}.
\end{aligned}
\end{equation}
Eqs. \ref{eq:T-EFIEe}, \ref{eq:T-EFIEi} are the {\it T-electric field integral equations} (T-EFIE), eqs. \ref{eq:T-MFIEi}, \ref{eq:T-MFIEe} are the 
{\it T-magnetic field integral equations} (T-MFIE). By subtracting eq. \ref{eq:T-EFIEi} from eq.  \ref{eq:T-EFIEe}  and eq.  
\ref{eq:T-MFIEi} from  \ref{eq:T-MFIEe} and substituting the expressions of the operators $\mathcal{E}_l^{\left(\Omega \right)}$
$\mathcal{H}_l^{\left(\Omega \right)}$, provided in the appendix \ref{sec:SurfaceSourceFields}, we obtain the PMCHWT formulation: 
\begin{equation}
 \mathcal{C}^{\left(\Omega \right)}{\bf x}^{\left(\Omega \right)} = {\bf y}^{\left(\Omega \right)},
\label{eq:GeneralT}
 \end{equation}
where we have defined the operator $\mathcal{C}^{\left(\Omega \right)}$
\begin{equation}
  \notag
  \mathcal{C}^{\left(\Omega \right)} = 
  {\left| {\begin{array}{*{40}c}
       \zeta_e \mathcal{T}_e^{ \left(\Omega,  t \right) }  &  + \mathcal{K}_e^{
\left(\Omega, t \right) } &   
     - \zeta_i  \mathcal{T}_i^{ \left(\Omega, t \right) }  &  - \mathcal{K}_i^{ \left(\Omega, t
\right) }   \\
       - \mathcal{K}_e^{ \left(\Omega,  t \right) }  &   {\zeta_e^{-1}}
\mathcal{T}_e^{\left(\Omega, t\right) }  & 
   +   \mathcal{K}_i^{ \left(\Omega,  t \right) } &   -{\zeta_i^{-1}}
\mathcal{T}_i^{\left(\Omega,
t\right) }  \\
\mathcal{I} & 0 & \mathcal{I} & 0 \\
0 & \mathcal{I} & 0 & \mathcal{I} 
\end{array}} \right|} ,
\end{equation}
the vector of unknowns ${\bf x}^{\left( \Omega\right)}$, and the excitation vector ${\bf y}^{\left( \Omega\right)}$
\begin{equation}
 \label{eq:PMCHWT_Excitation}
{\bf x}^{\left( \Omega\right)} = 
\left| {\begin{array}{*{20}c}
   {{\bf j}_e^{\left( \Omega, e \right)} }  \\
   {{\bf j}_e^{\left( \Omega, m \right)} }  \\
   {{\bf j}_i^{\left( \Omega, e \right)} }  \\
   {{\bf j}_i^{\left( \Omega, m \right)} } 
\end{array}} \right| ,
   \qquad
   {\bf y}^{\left( \Omega\right)} =  \left| {\begin{array}{*{20}c}
   { \frac{1}{2} {\bf n} \times {\boldsymbol\pi}_0^{ \left(\Omega,  m \right) }  }  \\
   {- \frac{1}{2} {\bf n} \times {\boldsymbol\pi}_0^{ \left(\Omega,  e \right) } }   \\
    {\boldsymbol\pi}_0^{ \left(\Omega,  e \right) } \\
    {\boldsymbol\pi}_0^{ \left(\Omega, m \right) }
\end{array}} \right|,
\notag
\end{equation}
and $\zeta_l \left\{ \Omega \right\}=\sqrt{{\mu_l}/{\varepsilon_l\left\{ \Omega \right\}}}$, with $l = i,e$.

\section{Numerical Solutions}
In the present section, we describe the numerical methods used to solve the PMCHWT formulation of eq.  \ref{eq:GeneralT} both at the
fundamental and at the SH frequency. In particular, we have exploited the Galerkin testing procedure to solve the SIEs, choosing the set of
basis functions to coincide with the set of test functions \cite{HarringtonBookMoM}.
\begin{figure}
\centering
\includegraphics[width=0.3\textwidth]{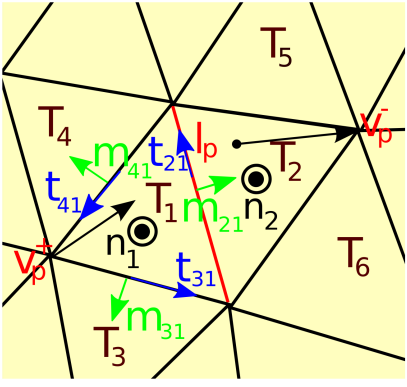}
\caption{The union of the two triangles  $T_1 = T_p^+,  T_2=T_p^-$ is the support of the basis function ${\bf f }_p$. For the triangle $T_p^+=T_1$ we
show the curves $l_{j1}$, their tangent vectors  ${\bf t}_{j1}$ and binormal vectors ${\bf m}_{j1}$ with j=2,3,4.}
  \label{fig:RWG}
\end{figure}

Let us introduce a triangular mesh with $N_t$ triangles and $N_e$ edges. We denote with $T_j$ the $j^{th}$ triangle, with $l_p$ the
length of the $p^{th}$ edge, with $\Sigma=\cup_j T_j$ the total surface of the mesh. We denote the two triangles sharing the $p^{th}$ edge as
$T_1=T_p^+$ and $T_2=T_p^-$ (Fig. \ref{fig:RWG}). They
have an area of $A_p^+$ and $A_p^-$ respectively.  The vertices of $T_p^+$ and $T_p^-$,  which are opposite to the $p^{th}$ edge, are indicated by 
$v_p^+$ and $v_p^-$ respectively. The unknown electric and magnetic currents  are expanded in terms of the Rao-Wilton-Glisson
(RWG) \cite{Rao82}  functions
set ${\bf f}_1, {\bf f}_2, \ldots, {\bf f}_{N_e}$. The RWG function relative to the $p^{th}$ edge is given by:
\begin{equation}
{\bf{f}}_p ({\bf{r}}) = \left\{ {\begin{array}{*{20}c}
  {\bf{f}}_p^+ ({\bf{r}}) = { + \frac{l_p}{2A_p^ +}\left( {{\bf{r}} - {\bf{v}}_p^ +  } \right)} & {{\bf{r}} \in T_p^ +  }  \\
  {\bf{f}}_p^- ({\bf{r}}) = { - \frac{l_p}{2A_p^ -  }\left( {{\bf{r}} - {\bf{v}}_p^ -  } \right)} & {{\bf{r}} \in T_p^ -  }  \\
   {\bf{0}} & \mbox{otherwise}  \\
\end{array}} \right.
\label{eq:RWG},
\end{equation}
and the corresponding support is $S_p = T_p^+ \cup T_p^-$ .
We introduce also the symmetric product  $ \langle {\bf f}, {\bf g} \rangle = \int_S {\bf f} \cdot {\bf g} dS $ for the space of square integrable
vector functions defined on S.

We  assembly the discrete operators associated to $\mathcal{K}_l^{\left( \Omega, t\right)}$ and $\mathcal{T}_l^{\left( \Omega, t\right)}$
 by projecting on the testing function ${\bf{f}}_p$  the image through the operators $\mathcal{K}_l^{\left( \Omega, t\right)}$, and
$\mathcal{T}_l^{\left( \Omega, t\right)}$ of the basis function ${\bf{f}}_q$.   Similarly, we assembly the discrete excitation vectors for both the
fundamental and SH problem by projecting on the testing function ${\bf{f}}_p$ the fields ${\boldsymbol\pi}_0^{\left( \omega, e \right)}$,
${\boldsymbol\pi}_0^{\left( \omega, m \right)}$,  ${\boldsymbol\pi}_0^{\left(2 \omega, e \right)}$,
${\boldsymbol\pi}_0^{\left( 2 \omega, m \right)}$. 

In order to compute the non linear sources we need to evaluate the fundamental field on the internal side of the surface $S$:
\begin{equation}
 \left.  {\bf E}_i^{\left( \omega \right)} \left( {\bf r} \right) \right|_{S_i}  = \left.{\bf E}_{i,t}^{\left( \omega \right)} \left( {\bf r} \right)
\right|_{S_i} + \left.{E}_{i,n}^{\left( \omega \right)} \left( {\bf r} \right) \right|_{S_i} \, {\bf n} ,
\notag
\end{equation}
where the tangent $ {\bf E}_{i,t}^{\left( \omega \right)}$ and normal component  ${E}_{i,n}^{\left( \omega \right)}$ of the electric field are:
\begin{equation}
  \begin{aligned}
     \left. {\bf E}_{i,t}^{\left( \omega \right)} \left( {\bf r} \right) \right|_{S_i} &= - {\bf n} \times {\bf j}_i^{\left( \omega, m
\right)}\left( {\bf r} \right), \\
 \left.{E}_{i,n}^{\left( \omega \right)} \left( {\bf r} \right) \right|_{S_i}  &= -j \frac{ \nabla_S \cdot {\bf j}_i^{\left(\omega, e
\right)}\left( {\bf r} \right)  }{\omega \varepsilon_i \left\{ \omega \right\}  }.
  \end{aligned}
\notag
  \end{equation}
Since the equivalent electric current ${\bf j}_i^{\left(\omega, e \right)}\left( {\bf r} \right) $ is represented through the linear RWGs functions,
$ {E}_{i,n}^{\left(\omega\right)} \left( {\bf r} \right)$ is a piecewise {\it constant} function presenting a discontinuity of the first
kind on the edges of the triangles. As a consequence,  $P_{n}^{S}$ (defined in 
eq.  \ref{eq:SurfacePolarization1}) is a piecewise constant function on the mesh and  the calculation
of the surface gradients $\nabla_S P_{n}^{S}$ needed to obtain $ {\boldsymbol\pi}_m^{\left( 2 \omega \right)}$ 
(see eq. \ref{eq:SurfaceMagneticCurrent_SH}) has to be carried out in the framework of the distribution theory.

We now introduce the following notations. For each triangle of the mesh, e.g. $T_1=T_p^+$ in Fig. \ref{fig:RWG}, we define
the curve $l_{j,1}$ having as support the edge of the triangle $T_1$ shared with the triangle $T_j$, which is oriented counterclockwise when
viewed from the end of the vector ${\bf n}_1$; ${\bf t}_{j1}$ is its tangent vector, and ${\bf m}_{j1} = {\bf t}_{j1} \times {\bf n}_1$ is
its binormal vector (orthogonal to both ${\bf t}_{j1}$ and $ {\bf n}_1$) (Fig. \ref{fig:RWG}).  Analogous quantities are also defined  for the
triangle $T_2=T_p^-$. Moreover, we denote with $\left. P_{n}^{S} \right|_{T_j} $ the value of the function $ P_{n}^{S}$ on the triangle $T_j$.
In order to simplify the treatment, we now assume that the triangles $\left\{ T_j, \quad j=1 \ldots 6 \right\}$ are coplanar. This hypothesis is
always true for particles of regular shapes and sufficiently dense meshes. 
The gradient of $P_{n}^{S}$ on the support of ${\bf f}_p$ can be written in the distribution sense as:
\begin{multline}
  \nabla_S {P}_{n}^{S} =
  \sum_{j=2,3,4}  \left(  \left. P_{n}^{S} \right|_{T_j} - \left. P_{n}^{S} \right|_{T_1} 
\right)   {\bf m}_{j1}  \delta_{j 1} + \\
  \sum_{j=5,6}  \left(  \left. P_{n}^{S} \right|_{T_j} - \left. P_{n}^{S} \right|_{T_2} 
\right)   {\bf m}_{j2}  \delta_{j2} + \\
  \label{eq:NablaPn}
\end{multline}
where the function $ \delta_{j 1}$ and  $ \delta_{j 2}$ are Dirac functions impulsive on the edge $l_{j1}$ and $l_{j2}$ respectively and 0 elsewhere
on $\Sigma$, such that $\iint_{\Sigma} \delta_{1j} dS = 1 $, $\iint_{\Sigma} \delta_{2j} dS = 1 $. In the appendix \ref{ExcitationVectors} we report
the explicit expressions of the discrete excitation vectors $ \langle {\bf f}_p,  {\boldsymbol \pi}_S^{\left( m \right)} \rangle$ and  $\langle {\bf
f}_p,  {\bf n} \times {\boldsymbol \pi}_S^{\left( m \right)} \rangle$.

Similarly, in order to compute the magnetic surface currents ${\boldsymbol\pi}_{\gamma'}^{\left( m \right)}$ introduced in eq.
\ref{eq:BulkMagneticCurrent_SHG}, we need to calculate the gradient of $P_{n}^{\gamma'}$. $P_{n}^{ \gamma' }$  is a piecewise quadratic
function presenting a discontinuity of the first kind along the mesh edges. Therefore, we obtain:

\begin{widetext}
\begin{multline}
  \nabla_S {P}_{n}^{\gamma'} = 
  \begin{cases}
  \displaystyle\sum_{j=2,3,4}  \left(  \left. P_{n}^{\gamma'} \right|_{T_j}\left( {\bf r} \right) - \left. {P}_{n}^{\gamma'}  \right|_{T_1} \left(
{\bf
r} \right) 
\right)  {\bf m}_{j1}  \delta_{j 1} +   \displaystyle\sum_{j=5,6}  \left(  \left. {P}_{n}^{\gamma'} \right|_{T_j}\left( {\bf r} \right) -
\left. {P}_{n}^{\gamma'}
\right|_{T_2} \left( {\bf r} \right)
\right)  {\bf m}_{j2}  \delta_{j2}  & \text{if $ {\bf r} \in   \partial{T}_1 \cup \partial{T}_2 $,} \\
 \nabla_S \left. {P}_{n}^{\gamma'}  \right|_{T_1} \left( {\bf r} \right) & \text{if $ {\bf r} \in   \overset{\circ}{T}_1 $,}
\\
 \nabla_S \left. {P}_{n}^{\gamma'}  \right|_{T_2} \left( {\bf r} \right) & \text{if $ {\bf r} \in  \overset{\circ}{T}_2 $,}
  \end{cases}
  \label{eq:NablaPb}
\end{multline} 
\end{widetext}
and $ \left. {P}_{n}^{\gamma'}  \right|_{T_j} \left( {\bf r} \right)$ is the restriction of the function ${P}_{n}^{\gamma'} \left( {\bf
r} \right)$ to the interior of the triangle $T_j$. In the appendix \ref{ExcitationVectors} we report the explicit expressions of the discrete
excitation vectors  $\langle {\bf f}_p,  {\boldsymbol \pi}_{\gamma'}^{\left( m \right)} \rangle$ and $\langle {\bf f}_p, {\bf n} \times {\boldsymbol
\pi}_{\gamma'}^{\left( m \right)} \rangle$.

Aiming at the evaluation of the excitation terms associated to ${\bf P}_{\delta'}$ we introduce a tetrahedral mesh of the particle volume $V_i$.
Then, we calculate the quantities ${\bf J}_{\delta'}$,  $\rho_{\delta'} $ and $\eta_{\delta'}$ in the quadrature points of the volumetric mesh using
the electric field ${\bf E}_i^{\left( \omega \right)}$ and the magnetic field ${\bf H}_i^{\left( \omega \right)}$ inside the particle at the
fundamental frequency. ${\bf E}_i^{\left( \omega \right)}$ and ${\bf H}_i^{\left( \omega \right)}$ are calculated using eq. \ref{eq:OperatorEH} in
combination with the methods described in \cite{Graglia93}. The derivation of the discrete excitation vectors  is straightforward in the Galerkin
formalism, and their expressions are reported in the appendix \ref{ExcitationVectors}.

\section{Discussion}
In the first part of this section, we validate our formulation for the problem of the SH generation from an isolated gold sphere using the full wave
SH Mie solution (SH-Mie) of the problems \ref{eq:FundamentalProblem} and \ref{eq:SHG_Problem}. The SH-Mie solution has been obtained by expanding
the pump field, the non-linear sources and the SH fields in series of vector spherical wave functions \cite{CaprettiPRB}. Then, we investigate the
SH generation properties of a triangular nano-prism. The
scatterers are homogeneous gold particles, embedded in air. The investigated  physical quantity is the SH radiated power per unit solid angle
collected along the direction $\bf k$
\begin{equation}
   {d P^{\left( \Omega \right) }}/{d\Omega} \left( {\bf k} \right) =\lim_{ \left\| {\bf r} \right\| \to +\infty} \left[ \frac{r^2}{2
\zeta_e\left\{ \Omega \right\}} \left\| {\bf E}_e^{\left( \Omega \right)} \left( {\bf k}\right) \right\|^2 \right].
\end{equation}
The gold dispersion is modeled by fitting experimental  data \cite{Johnson1972}.  The integrals involved in the numerical solution have been evaluated
using four Gauss quadrature points for each triangle, and their singularities have been managed exploiting singularity extraction techniques
\cite{Graglia93}. The volume integrals have been evaluated using a single Gauss quadrature point for each tetrahedron.

The gold sphere is excited by a plane wave of unitary intensity, namely ${\bf E}_0^{\left( \omega \right)}=\exp \left( -j \omega \sqrt{\varepsilon_e
\mu_e}z\right) \hat{\bf x}$ propagating along the positive direction of the $z$-axis and linearly polarized along
$x$. The exciting wavelength corresponds to the plasmonic resonance of a gold spherical particle in the Rayleigh limit, {\it i.e.} $520nm$. The
relative
permittivity used
in the calculations are $\varepsilon_i \left ( 520nm \right) = -3.88 - 2.63 j $ and $\varepsilon_i \left (260 nm \right) = -1.20 - 4.67j $ for the
fundamental and the SH problem, respectively. From both theoretical and experimental studies it results that the component ${\chi}_{\perp \parallel 
\parallel}^{\left( 2 \right)} $ only weakly contributes to the SH generation from noble metals \cite{Bachelier10}. Moreover, in the case of the sphere
following Bachelier et al. \cite{Bachelier10} we also neglect the contribution of the $\delta'$ source.
 Following \cite{Sipe80,Bachelier10} we express the parameters  ${\chi}_{\perp \perp \perp}$, ${\chi}_{\parallel \perp \parallel}$, and
$\gamma $ through the dimensionless {\it Rudnick-Stern parameters} $a$, $b$  and $d$ \cite{Sipe80} as follows:
\begin{equation}
    \label{eq:Sipe}
     \left[ {\chi}_{\perp \perp \perp}^{\left( 2 \right)},  {\chi}_{\parallel \perp \parallel}^{\left( 2 \right)}, \gamma \right]
= -\left[\frac{a}{4}, \frac{b}{2},  \frac{d}{8} \right]  \chi_b^{\left( 1 \right)} \left( \omega \right)
\frac{\omega_p^2}{\omega^2} \frac{\varepsilon_0}{e n_0},
\end{equation}
where $ \chi_b^{\left( 1 \right)} \left( \omega \right) =\left( \epsilon_i \left( \omega \right)/\epsilon_0 - 1\right)$  is the bulk linear
susceptibility of the metal, $e$ is the electron charge (absolute value), $n_0$ is the number density of the conduction electrons, $\omega_p$ is the
plasma frequency and $\varepsilon_0$ is the permittivity of the vacuum. By choosing $a=1$, $b=-1$  and $d=1$ we obtain the Rudnick-Stern
hydrodynamic model \cite{Rudnick71,Sipe80}.
 The mesh has a number of edges equal to $N_e=3747$.  The maximum  expansion order of 
vector spherical wave functions, used in the calculation of the SH-Mie solution, is   $N=10$.
\begin{figure*}
  \centering
  \includegraphics[width=\textwidth]{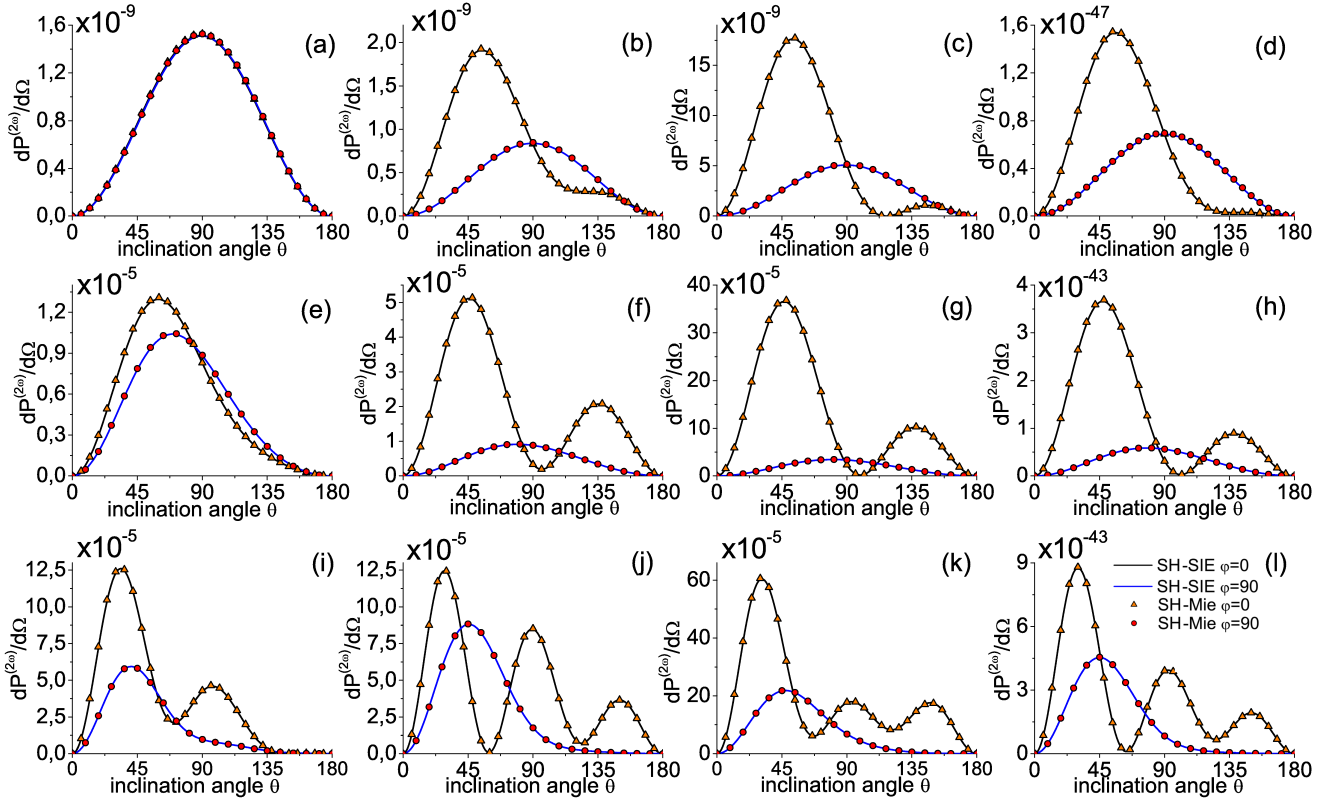}
  \caption{SH radiated power per unit solid angle (expressed in Watt/Steradian) of a gold sphere with diameter (a-d) 20nm, (e-h) 100nm (i-l)
200nm. Different combinations of SHG polarization sources have been considered, namely $\left( \gamma, {\chi}_{\parallel \perp
\parallel}^{\left( 2 \right)}, {\chi}_{\perp \perp \perp}^{\left( 2 \right)} \right) = $ $\left( 1, 0, 0\right)$ in panels (a,e,i); $\left( 0, 1,
0\right)$ in panels
(b,f,j); $\left( 0, 0, 1\right)$ in panels (c,g,k).  In panels (d,h,l) the sources have been weighted using eq. \ref{eq:Sipe} with
$\left(a,b,d\right)=\left(1,-1,1\right)$. The sphere is excited by a x-polarized plane wave of unitary intensity propagating along 
the positive $z$ axis with wavelength 520nm. The diagrams are in linear scale.}
  \label{fig:ComparisonSIE_MIE}
\end{figure*}
Aiming at an accurate validation of the proposed approach, we show in fig. \ref{fig:ComparisonSIE_MIE} the quantity $ {d P^{\left( 2\omega \right)
}}/{d\Omega}$ as a function of the inclination angle $\theta$ for $\phi=0$ and $\phi=90$. These quantities have been calculated by the SH-Mie (orange
triangles for $\phi=0$ and red circles for  $\phi=90$) and by the SH-SIE (continuous black line for $\phi=0$ and continuous blue line for  $\phi=90$).
Several sphere's diameters $D$ have been considered: the first row of fig. \ref{fig:ComparisonSIE_MIE} (panels a-d), is relative to an electrically
small particle with $D=20nm$, the second row (e-h) to $D=100nm$, the third row (i-l) to $D=200nm$. We considered in the first three columns only one
SH source at a time, whereas in the fourth column the different nonlinear polarization sources are simultaneously active. In particular, in the first
column of fig. \ref{fig:ComparisonSIE_MIE}, corresponding to panels (a,e,i), we considered only the nonlocal-bulk source $P_{n}^{\gamma'}$ using the
coefficients $\left( \gamma, {\chi}_{\parallel \perp \parallel}^{\left( 2 \right)}, {\chi}_{\perp \perp \perp}^{\left( 2 \right)}
\right) = $ $\left( 1, 0, 0\right)$. In the second column (b,f,j) only the surface source $P_{t}^{S}$ has been considered
assuming $\left( \gamma, {\chi}_{\parallel \perp \parallel}^{\left( 2 \right)}, {\chi}_{\perp \perp \perp}^{\left( 2 \right)} \right) =\left( 0, 1,
0\right)$, whereas in the third column (c,g,k) only the surface source $P_{n}^{S}$ has been considered assuming $\left(
\gamma, {\chi}_{\parallel \perp \parallel}^{\left( 2 \right)}, {\chi}_{\perp \perp \perp}^{\left( 2 \right)} \right) =\left( 0, 0, 1\right)$.
Eventually, in panels (d,h,l,p) we considered a realistic physical case in which the different weights are calculated using eq. \ref{eq:Sipe} with  
$\left(a,b,d\right)=\left(1,-1,1\right)$. In all the different cases shown in fig. \ref{fig:ComparisonSIE_MIE} very good agreement between the
SH-SIE and the SH-Mie formulations can be found.

\begin{figure}
  \centering
  \includegraphics[width=0.5\textwidth]{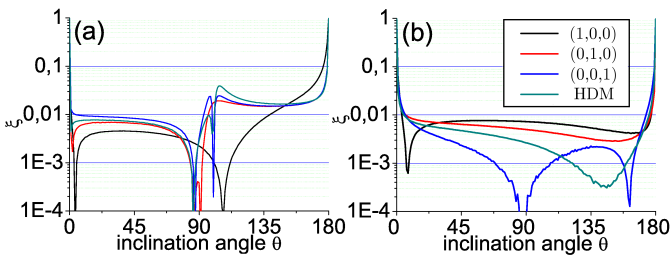}
  \caption{Relative error of ${d P^{\left( 2\omega \right) }}/{d\Omega}$ calculated with the SH-SIE with respect to the SH-Mie solution for a
sphere of diameter $D=100nm$ as a function of the inclination angle $\theta$ for (a) $\phi=0$ and (b) $\phi = 90$.  Different combinations of SHG
polarization sources have been considered, namely $\left( \gamma, {\chi}_{\parallel \perp \parallel}^{\left( 2 \right)},
{\chi}_{\perp \perp \perp}^{\left( 2 \right)} \right) = $ $\left( 1, 0,0\right)$ (black line); $\left( 0, 1, 0\right)$ (red line); $\left( 0, 0,
1 \right)$ (blue line); the hydrodynamic model of eq. \ref{eq:Sipe} with  $\left(a,b,d\right)=\left(1,-1,1\right)$ (green line).}
  \label{fig:ErrorSIE}
\end{figure}

 In Fig. \ref{fig:ErrorSIE} we calculate the relative error $\xi$ of the SH-SIE formulation with respect to the SH-Mie formulation, defined as:
\begin{equation}
   \xi =  \left(      {d P^{\left( 2\omega \right) }}/{d\Omega} - {P_\texttt{MIE}^{\left( 2\omega
\right)}}/{d\Omega}\right)/ \left({{P^{\left( 2\omega \right)}}/{d\Omega}}\right),
\end{equation}
where  $  {P}_\texttt{MIE}^{\left( 2\omega \right) }/{d\Omega}$ is the radiated power per unit solid angle calculate with the SH-Mie formulation.
The error is plotted  as a function of inclination angle $\theta$ for (a) $\phi=0$ and (b) $\phi = 90$ for  of the for the $D=100nm$ sphere.
All the investigated cases present a comparable error,  below $3 \%$. Only in correspondence of the forward and backward scattering
direction, when the corresponding value of  ${d P^{\left( 2\omega \right) }}/{d\Omega}$ is very small the error is higher. 

\begin{figure}
  \centering
  \includegraphics[width=0.5\textwidth]{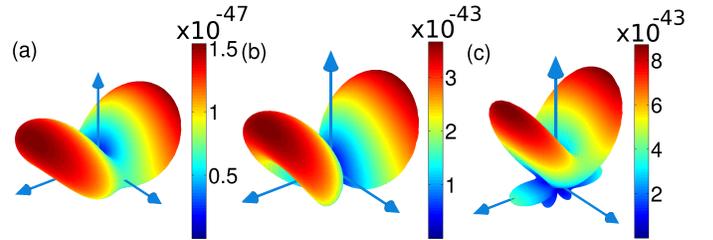}
  \caption{Radiation diagram of  $ {d P^{\left( 2\omega \right) }}/{d\Omega}$ (expressed in Watt/Steradian) for a gold sphere with diameter: (a)
20nm, (b) 100nm (c) 200nm. The sphere is excited by a x-polarized plane wave of unitary intensity propagating along the positive direction
of the z-axis with wavelength 520nm. The radiated power is evaluated at 260nm. The diagrams are in linear scale. The sources have been
weighted using eq. \ref{eq:Sipe} with  $\left(a,b,d\right)=\left(1,-1,1\right)$.}
  \label{fig:dSCS_Sphere}
\end{figure}


In figure \ref{fig:dSCS_Sphere} we show the  radiation diagram of  $ {d P^{\left( 2\omega \right) }}/{d\Omega}$ for different diameters of the
sphere, $D=$ (a) $20nm$ (b) $100nm$ (c) $200nm$. The SH radiated power vanishes for the forward and backscattering directions for all the
investigated diameters,  because the sphere is rotationally symmetric around the propagation direction. As we increase the sphere's diameter we notice
the appearance of multiple oscillations corresponding to high order modes. Moreover, as the particle size increases, the inclination angle
corresponding to the first maximum SH intensity approaches the forward scattering direction. This behavior has been experimentally demonstrated for
silver spherical particles in a recent work by Gonella {\it et al.} \cite{Gonella12}.

We now consider a triangular nano-prism of height $h=40nm$ with rounded edges of length $200nm$. The radius of the osculating cylinder (sphere) at
each edge (corner) is $10nm$. This shape is of extreme interest in nanoplasmonics as the building block of bow-tie nano-antennas, whose SH
properties have been recently investigated in \cite{Hanke09}. We chose as the exciting wavelength $\lambda=690nm$, corresponding to the peak
 of the extinction cross section over the visible range.
\begin{figure}
  \centering
  \includegraphics[width=0.5\textwidth]{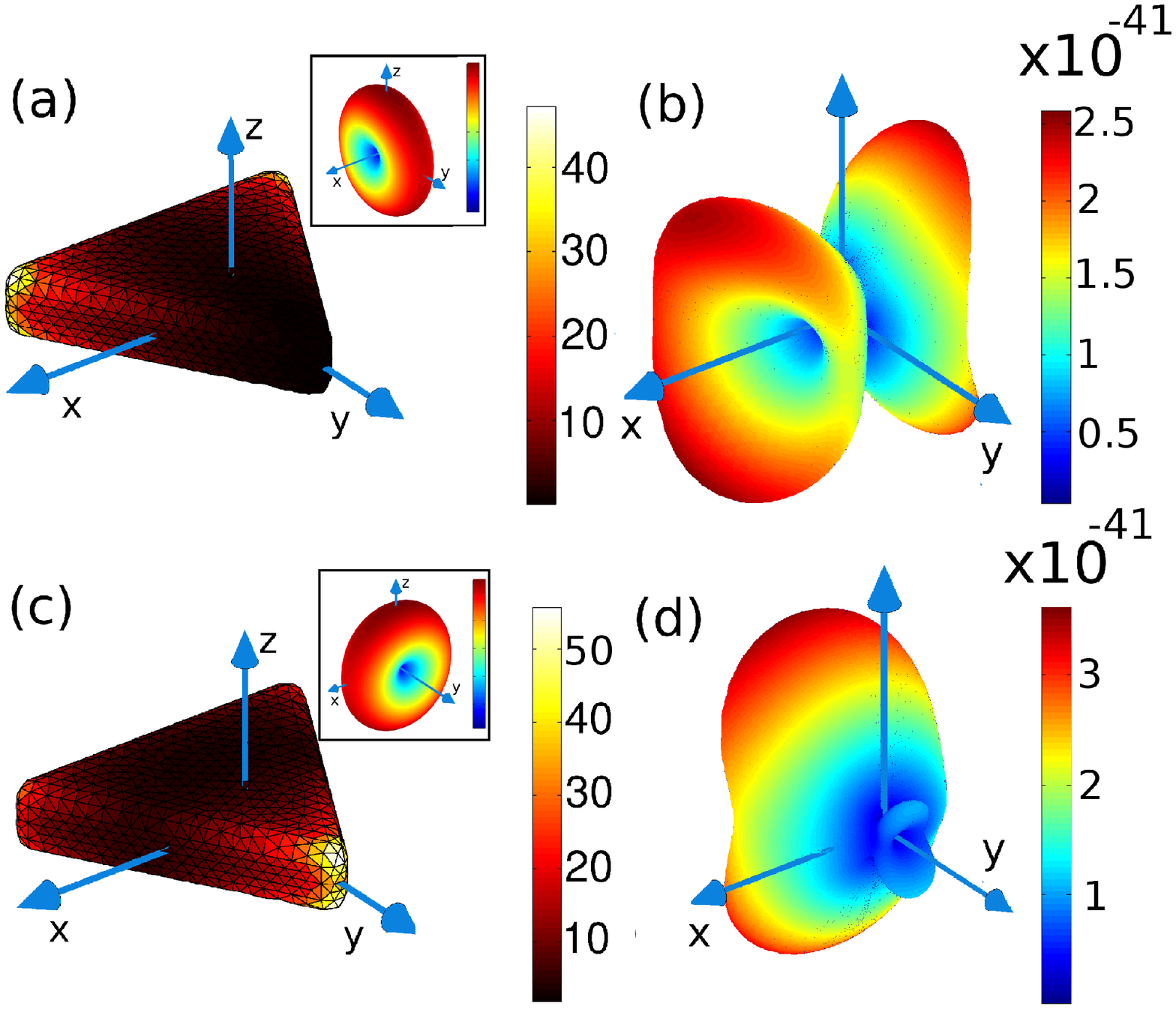}
  \caption{(a,c) Magnitude of the field ${\bf E}_e^{\left( \omega \right)}$ at the fundamental frequency  on the surface of the triangular
nano-prism excited by a monochromatic ($\lambda=690nm$) plane wave of unitary intensity propagating along the positive $z-axis$ and linearly polarized
along (a) $x$ and (c) $y$, and the corresponding radiated power per unit solid angle (inset). (b,d) Radiation diagram of  $ {d P^{\left( 2\omega
\right) }}/{d\Omega}$ (expressed in Watt/Steradian) for both the (b) x-polarized and (d) y-polarized pump.}
  \label{fig:TriPrism_Fund}
\end{figure}
Due to the irregular shape the distribution of the field will be highly inhomogeneous and the relaxation term of the free-electron in the hydrodynamic
model has to be considered \cite{Benedetti10, CentiniSPIE, Benedetti11, Wang09}. As a consequence, we correct the Rudnick-Stern coefficients
following ref. \cite{Benedetti11}. Moreover, the contribution due to $\delta'$ is non-vanishing; it is related to $\gamma$ by
\cite{CentiniSPIE, Benedetti11, Wang09}:
\begin{equation}
       \delta' =- \frac{2}{1+ j \omega \tau} \gamma
      \notag
\end{equation}
where $\tau = 9.3\cdot 10^{-15}$ \cite{Johnson1972}.
We first analyze the scattering properties at the fundamental frequency. In figure \ref{fig:TriPrism_Fund} (a,c) we show the magnitude
of the total electric field at the fundamental frequency on the surface of the triangular nano-prism when the incident plane wave is
polarized along (a) the $x$ axis and (c) the $y$ axis. When the incident plane wave is x-polarized two {\it hot spots} arise at the two ends
of the edge parallel to the incident polarization, whereas in correspondence of a y-polarized excitation the remaining corner experiences the highest
field enhancement. The radiated power  ${dP}^{\left( \omega \right)}/ d\Omega$ at the fundamental frequency can be attributed to an electric
dipole oriented along the polarization direction (inset of Fig. \ref{fig:TriPrism_Fund} (a,c)). In panels (b,d) we study the SH radiated power 
$ {d P^{\left( 2\omega \right) }}/{d\Omega}$ by the triangular nano-prism.
Unlike the linear scattering, $ d P^{\left( 2 \omega \right)} / d \Omega $ shows a richer behavior dominated by higher order modes. Furthermore, the
shape of $ d P^{\left( 2 \omega \right)} / d \Omega $ is affected by the
distribution of the electric field on the surface of the particle, shown in fig. \ref{fig:TriPrism_Fund} (a,b). In fact, we attribute the presence of
the two lobes in panel to the strong field localization that occurs in correspondence of the two corners of the edge of the triangular-nanoprism
parallel to x-axis. In contrast, when the incident field is directed along the y-axis, the radiation diagram features one lobe driven by the hot-spot
on the remaining corner. Although both the surface area and the volume of the nano-prims are lower than those of the sphere of diameter $200nm$
investigated earlier, the maximum intensity of SH radiated power per unit solid angle $ d P^{\left( 2 \omega \right)} / d \Omega $ of the nano-prism
(Fig. \ref{fig:TriPrism_Fund}(b,d))  is almost two order of magnitude higher compared to the sphere (Fig. \ref{fig:dSCS_Sphere} (b)).  This fact is
due to both the higher values of the non-linear susceptibilities of gold at $690nm$ and to the NP asymmetric shape
\cite{Canfield07}.

\section{Conclusions}
We have presented a surface integral equation method to  solve the second harmonic generation problem in centrosymmetric nanoparticles with
arbitrary shapes accounting for both the nonlocal-bulk and local-surface nonlinear polarization sources. We have demonstrated that the contribution 
of the nonlocal-bulk nonlinear polarization sources to the SH electromagnetic field can be taken into account by introducing equivalent
electric and magnetic currents on the nanoparticle boundary. We compared our numerical solution to the analytical Mie solution for spheres of several
radii, obtaining very good agreement for any investigated case. The developed method paves the way to a better understanding of the process of
SHG in arbitrarily shaped nanoparticles. This approach can have a high impact in the design of novel nanoplasmonic devices with enhanced SHG 
emission, including for instance sensor probing physical and chemical properties of material surfaces.
\section{Acknowledgement}
This work was partly supported by the Italian Ministry of Education, University and Research through the project PON01\_02782.
\appendix
\section{Electromagnetic field generated by surface sources}
\label{sec:SurfaceSourceFields}
In the present section, we provide the expression of the electromagnetic field generated by a surface distribution of electric and magnetic currents 
$ \left( {\bf j}^{\left(\Omega, e \right)}, {\bf j}^{\left(\Omega, m \right)} \right)$  defined on $S$ and radiating in a homogeneous medium with
constitutive parameters $ \left( \varepsilon_l \left\{ \Omega \right\}, \mu_l \right) $ at frequency $\Omega$.   The electromagnetic field radiated by
both the electric and magnetic magnetic currents is \cite{VanBladel}:
\begin{subequations}
  \begin{align}
    {\bf{E}}^{\left( \Omega \right)} \left( {\bf{r}} \right) &= \mathcal{E}_l^{\left( \Omega \right)} \left\{ {{\bf j}^{\left( \Omega, e \right)}
,{\bf j}^{\left(  \Omega,  m \right)} } \right\}\left(
{\bf{r}}
\right) \\
    {\bf{H}}^{\left( \Omega \right)} \left( {\bf{r}} \right) &= \mathcal{H}_l^{\left( \Omega \right)} \left\{ {{\bf j}^{\left(  \Omega,  e \right)}
,{\bf j}^{\left( \Omega,  m \right)} } \right\}\left(
{\bf{r}}
\right) 
  \end{align}
\end{subequations}
where: 
\begin{widetext}
   \begin{equation}
      \begin{aligned}
      \mathcal{E}_l^{\left( \Omega \right)} \left\{ {\bf j}^{\left(  \Omega,  e \right)}, {\bf j}^{\left(  \Omega, m \right)} \right\} &= \zeta_l
\left\{ \Omega \right\} {\mathcal{T}}_l^{\left(  \Omega \right)}  \left\{ {{\bf j}^{\left(  \Omega, e \right)} } \right\}
                + \mathcal{K}_l^{\left(  \Omega \right)} \left\{{{\bf j}^{\left(  \Omega, m \right)} } \right\}+ 
 \begin{cases}
    {\bf{0}} & \text{if $ {\bf r} \notin S$,} \\
+ \frac{\zeta_l \left\{ \Omega \right\} }{2jk_l \left\{ \Omega \right\} } \left[ {\nabla_S \cdot {\bf j}^{\left( \Omega,  e \right)}}\right] {\bf{n}} 
-  \frac{1}{2}{\bf{n}} \times {{\bf j}^{\left(  \Omega, m \right)} } & \text{if $ {\bf r} \in  S_{i}$,} \\
- \frac{\zeta_l \left\{ \Omega \right\}}{2jk_l \left\{ \Omega \right\}} \left[ {\nabla_S \cdot {\bf j}^{\left(  \Omega, e \right)} } \right]{\bf{n}} 
  +  \frac{1}{2}{\bf{n}} \times{{\bf j}^{\left(  \Omega, m \right)} }  & \text{if ${\bf r} \in  S_{e}$,}
\end{cases} \\
   \mathcal{H}_l^{\left( \Omega \right)} \left\{ {{\bf j}^{\left(  \Omega, e \right)}, {\bf j}^{\left(  \Omega, m \right)} } \right\} &= 
 - \mathcal{K}_l^{\left(  \Omega \right)} \left\{ {{\bf j}^{\left(  \Omega, e \right)} } \right\}+
   \frac{\mathcal{T}_l^{\left(  \Omega \right)} \left\{ {\bf j}^{\left( \Omega, m\right)} \right\}  }{\zeta_l \left\{ \Omega \right\} } 
   + 
 \begin{cases}
    {\bf{0}}  & \text{if $ {\bf r} \notin S$,} \\
  + \frac{1}{2}{\bf{n}} \times{{\bf j}^{\left(  \Omega, e \right)} \left( {\bf{r} } \right)} 
  + \frac{\left[{\nabla_S \cdot {\bf j}^{\left( \Omega,  m \right)} \left( {\bf{r} }
\right)}\right]}{2j\zeta_l \left\{ \Omega \right\} k_l \left\{ \Omega \right\}}  {\bf{n}} & \text{if $ {\bf r} \in  S_{i}$,} \\
-\frac{1}{2}{\bf{n}}\times{{\bf j}^{\left(  \Omega, e \right)} \left( {\bf{r} } \right)}
- \frac{\left[{\nabla_S \cdot {\bf j}^{\left(  \Omega, m \right)} \left( {\bf{r} }
\right)}\right]}{2j\zeta_l \left\{ \Omega \right\} k_l\left\{ \Omega \right\}}{\bf{n}} & \text{if ${\bf r} \in 
S_{e}$,}
\end{cases}
\label{eq:OperatorEH}
      \end{aligned}
   \end{equation}
\end{widetext}

 $\nabla_S \cdot$ denotes the surface divergence, $\mathcal{K}_l$ and $\mathcal{T}_l$ denote the integral operators: 
\begin{subequations}
  \begin{align}
    \label{eq:Operator_K}
    \mathcal{K}_l^{\left( \Omega\right)} \left\{ {\bf w} \right\} \left( {\bf r} \right) &= - \int_{S} {\bf w}  \left( {\bf r}' \right) \times
\nabla' g_l^{\left( \Omega\right)} \left( {\bf r}  - {\bf r}' \right) dS'  \\
    \label{eq:Operator_T}
    \mathcal{T}_l^{\left( \Omega\right)} \left\{ {\bf w} \right\} \left( {\bf r} \right) &=  - jk_l \left\{ \Omega \right\} \int_S g_l^{\left(
\Omega\right)} \left( {\bf r} - {\bf r}' \right) {\bf w} \left( {\bf r}'
\right) dS'  + \notag \\
& - \frac{1}{jk_l\left\{ \Omega \right\} } \int_{S} \nabla' g_l ^{\left( \Omega\right)} \left( {\bf r}  - {\bf r'} \right) \nabla' _S 
\cdot {\bf w}\left( {\bf r}' \right) dS' 
  \end{align}
\end{subequations}
$g_l^{\left( \Omega\right)}$ is the homogeneous space Green's function, {\it i.e.}
$   g_l^{\left( \Omega\right)}\left( {\bf r} - {\bf r}' \right) = \frac{ e^{- j k_l \left\{ \Omega \right\} \left|{\bf r} - {\bf r}' \right|} }{4 \pi
\left|{\bf r} - {\bf r}' \right| } $, $k_l \left\{ \Omega \right\}=\Omega \sqrt{\mu_l \varepsilon_l \left\{ \Omega \right\} }$.

\begin{subequations}
\begin{align}
 \label{eq:Kt_Continuous}
  \mathcal{K}_l^{\left(\Omega,  t \right)} \left\{\cdot  \right\} &= \left. -{\bf n} \times  {\bf n} \times \mathcal{K}_l^{ \left( \Omega
\right) } \left\{ \cdot \right\}
\right|_{S_l} \\
  \label{eq:Tt_Continuous}
  \mathcal{T}_l^{\left(\Omega, t \right)} \left\{\cdot  \right\} &= \left. -{\bf n} \times  {\bf n} \times \mathcal{T}_l^{ \left(
\Omega
\right) } \left\{ \cdot  \right\}
\right|_{S_l}.
\end{align}
\end{subequations}

\section{Excitation Vectors}
\label{ExcitationVectors}
Now we calculate the discrete excitation vector associated to the superficial vector field $ {\boldsymbol \pi}_m^{\left( 2 \omega
\right)}$:
\begin{equation}
  \langle {\bf f}_p,  {\boldsymbol \pi}_S^{\left( m \right)} \rangle = \iint_{\Sigma} {\bf f}_p    
\cdot {\boldsymbol \pi}_S^{\left( m \right)} dS 
\end{equation}
Using eqs. \ref{eq:SurfaceMagneticCurrent_SH} and \ref{eq:NablaPn} we obtain:
\begin{multline}
   \langle {\bf f}_p, {\boldsymbol \pi}_S^{\left( m \right)} \rangle  =   \\
\frac{1}{ \varepsilon_0} \sum\limits_{j = 2,3,4} \iint_\Sigma \left( \left.  P_{n}^S
\right|_{T_j}  - \left. P_{n}^S \right|_{T_1}  \right)  {\bf f}_p \cdot {\bf n} \times   {\bf m}_{j1}  \delta_{j1} dS  \\
+\frac{1}{ \varepsilon_0} \sum\limits_{j = 5,6} \iint_\Sigma \left( \left.  P_{n}^S
\right|_{T_j}  - \left. P_{n}^S \right|_{T_2}  \right)  {\bf f}_p \cdot {\bf n} \times  {\bf m}_{j2}  \delta_{j2} dS 
  \label{eq:Normal_t_StepII}
\end{multline}
It is worth noting that the function ${\bf f}_p$ and ${\bf n}$ is discontinuous in $\Sigma$. Therefore, we have to use the following properties of
the Dirac delta function. Let ${\bf F}$ be a function defined as ${\bf F} \left( {\bf r} \right) =  \left. {\bf F} \right|_{T_{1}}\left( {\bf r}
\right)  \, \mbox{ if } \, {\bf r} \in {T_j}$ where each function $\left. {\bf F} \right|_{T_{1}}$ is regular in the interior of the triangle $T_j$.
The following relation holds
in the
distribution sense:
\begin{equation}
   {\bf F} \delta_{l_{j1}} = \frac{ \left. {\bf F} \right|_{T_{j}} + \left. {\bf F} \right|_{T_{1}}}{2} \delta_{l_{j1}}
\end{equation}
Applying this property to the integrands of eq. \ref{eq:Normal_t_StepII} and exploiting the fact that ${\bf f}_p=0$ in $T_{3},T_{4},T_{5},T_{6}$,
${\bf f}_p$ = ${\bf f}_p^+$ in $T_1$
and ${\bf f}_p$ = ${\bf f}_p^-$ in $T_2$ we obtain:
\begin{multline}
    {\bf f}_p \cdot {\bf n} \times  {\bf m}_{j1}  \delta_{j1} = \\
   \begin{cases}
     \frac{1}{2}  {\bf f}_p^+ \cdot {\bf n}_1 \times  {\bf
m}_{j1} 
\delta_{j1} & \text{j=3,4} \\
     \frac{1}{2}  \left( {\bf f}_p^+ \cdot {\bf n}_1 \times  
{\bf m}_{j1}  + {\bf f}_p^- \cdot {\bf n}_2 \times   {\bf m}_{j1} \right) \delta_{j1}
& \text{j=2}
   \end{cases}
\notag
\end{multline}
\begin{equation}
   {\bf f}_p \cdot {\bf n} \times  {\bf m}_{j2}  \delta_{j2} = 
    \left(  {\bf f}_p^- \cdot {\bf n}_2 \times   {\bf m}_{j2} \right) \delta_{j2} \qquad \text{j=5,6}
\notag
\end{equation}
Therefore, using the sampling property of the Dirac delta function eq. \ref{eq:Normal_t_StepII} becomes:
\begin{multline}
     \langle {\bf f}_p, {\boldsymbol \pi}_m^{\left( S \right)} \rangle = 
    \frac{1}{2 \varepsilon_0} \sum_{j=2,3,4} \left( \left.  P_{n}^S \right|_{T_j}  - \left. P_{n}^S \right|_{T_1}  \right) 
  \int_{l_{j1}} {\bf f}_p^+ \cdot  {\bf t}_{j1} dl \\
  + \frac{1}{2 \varepsilon_0} \sum_{j=1,5,6} \left( \left.  P_{n}^S \right|_{T_j}  - \left. P_{n}^S \right|_{T_2}  \right) 
 \int_{l_{j2}} {\bf f}_p^- \cdot  {\bf t}_{j2} dl 
\notag
\end{multline}
Moreover, since we have   $\oint_{\partial T_1} {\bf f}_p^+ \cdot   d{\bf l} = 0$ and $\oint_{\partial T_2} {\bf f}_p^- \cdot   d{\bf l} = 0$ 
we finally obtain:
\begin{multline}
     \langle {\bf f}_p, {\boldsymbol \pi}_m^{\left( S \right)} \rangle = 
    \frac{1}{2 \varepsilon_0} \sum_{j=2,3,4}  \left.  P_{n}^S \right|_{T_j}    
  \int_{l_{j1}} {\bf f}_p^+ \cdot  {\bf t}_{j1} dl \\
  + \frac{1}{2 \varepsilon_0} \sum_{j=1,5,6} \left.  P_{n}^S \right|_{T_j}  
 \int_{l_{j2}} {\bf f}_p^- \cdot  {\bf t}_{j2} dl 
\notag
\end{multline}

In conclusion, in order to compute the term $ \langle {\bf f}_p, {\boldsymbol \pi}_S^{\left( m \right)} \rangle $ we have to consider not only the
value of $P_n^S$ in the triangles $T_p^+$ and $T_p^-$ but also in the 4 adjacent triangles that have one edge in common with $T_p^+$ and $T_p^-$.

Using a similar derivation and exploiting the fact that ${\bf f}_p^+ \cdot {\bf n}_1 \times {\bf n}_1 \times  {\bf m}_{j1} =
{\bf 0} $ with $j=3,4$ and ${\bf f}_p^- \cdot {\bf n}_2  \times {\bf n}_2  \times {\bf m}_{j2}=0$ with $j=5,6$ it is possible to
prove that:
\begin{multline}
   \langle {\bf f}_p,  {\bf n} \times {\boldsymbol \pi}_S^{\left( m \right)} \rangle = \\   - \frac{1}{2 \varepsilon_0} \left( \left.  P_{n}^S
\right|_{T_1} -  \left. P_{n}^S \right|_{T_2}  \right)   {\bf m}_{12}  \cdot  \int_{l_{12}} \left(  {\bf f}_p^+  + {\bf f}_p^-  \right) dl = \\
   \frac{l_p}{\varepsilon_0} \left( \left.  P_{n}^S
\right|_{T_1} -  \left. P_{n}^S \right|_{T_2}  \right)   
\notag
\end{multline}
The treatment of the excitation terms arising from the  bulk source $P_{n}^{\gamma'}$ follows the same steps of  $P_{n}^S$
\begin{multline}
     \langle {\bf f}_p,  {\boldsymbol \pi}_{\gamma'}^{\left( m \right)} \rangle = \\
    \frac{1}{2 \varepsilon_0} \sum_{j=2,3,4}  \int_{l_{j1}}  \left( \left.  P_{n}^{\gamma'} \right|_{T_j}  + \left. P_{n}^{\gamma'} 
\right|_{T_1} \right) {\bf f}_p^+ \cdot {\bf
t}_{j1}  dl \\
  + \frac{1}{2 \varepsilon_0} \sum_{j=1,5,6}  \int_{l_{j2}} \left( \left. 
P_{n}^{\gamma'} \right|_{T_j}  + \left. P_{n}^{\gamma'}  \right|_{T_2} \right) {\bf f}_p^-  \cdot   {\bf t}_{j2}  dl 
\notag
\end{multline}
Using similar arguments it is possible to demonstrate that:
\begin{multline}
   \langle {\bf f}_p, {\bf n} \times {\boldsymbol \pi}_{\gamma'}^{\left( m \right)} \rangle = \\
   \frac{1}{\varepsilon_0}  \iint_{T_1} \nabla_S \cdot {\bf f}_p^+   \left. P_{n}^{\gamma'} \right|_{T_1}  \, dS + \frac{1}{\varepsilon_0} 
\iint_{T_2} \nabla_S \cdot {\bf f}_p^- \left. P_{n}^{\gamma'}  \right|_{T_2}  \, dS 
\notag
\end{multline}
In order to evaluate the excitation terms relative to ${\bf P}_{\delta'}$ we have to perform volumetric integrals and therefore we introduce a 
tetrahedral mesh of the particle volume $V_i$. The derivation of the corresponding source terms is straightforward in the Galerkin formalism.
Using eqs. \ref{eq:BulkMagneticCurrent_SHG} and \ref{eq:E_Delta1} we have:
\begin{multline}
  \langle {\bf f}_p  , {\boldsymbol\pi}_{\delta'}^{\left( m \right)} \rangle = \sum_{r=\pm} \Biggl\{ \\
 -j \omega \mu_0  \iint_{T_p^r} {\bf n} \times {\bf f}_p \left( {\bf r} \right) \cdot \left[  \iiint_{V_i}
{\bf J}_\delta \left( {\bf r}' \right) g_i \left( {\bf r } -{\bf r}' \right) dV' \right] dS \\
  + \frac{1}{\varepsilon_i \left\{ 2 \omega \right\} } \iiint_{V_i} \rho_\delta \left( { \bf r}' \right)  \left[ \oint_{\partial T_p^r} g_i
\left( {\bf r } -{\bf r}' \right)  {\bf f}_p \left( {\bf r} \right) \cdot d{\bf l} \right] dV' \\
  + \left. \frac{1}{\varepsilon_i \left\{ 2 \omega \right\} } \iint_{T_p^r} \eta_\delta \left( { \bf r}' \right)  \left[ \oint_{\partial T_p^r} g_i
\left( {\bf r } -{\bf r}' \right)  {\bf f}_p \left( {\bf r} \right) \cdot d{\bf l} \right] dS' \right\} 
\notag
\end{multline}
\begin{multline}
  \langle {\bf f}_p  , {\bf n} \times {\boldsymbol\pi}_{\delta'}^{\left( m \right)} \rangle = \sum_{r=\pm} \Biggl\{ \\
 -j \omega \mu_0  \iint_{T_p^r} {\bf f}_p \left( {\bf r} \right) \cdot \left[  \iiint_{V_i}
{\bf J}_\delta \left( {\bf r} \right) g_i \left( {\bf r } -{\bf r}' \right) dV' \right] dS \\
 +  \frac{1}{\varepsilon_i \left\{ 2 \omega \right\} } \iiint_{V_i} \rho_\delta \left( { \bf r}' \right)  \left[ \iint_{ T_p^r} g_i
\left( {\bf r } -{\bf r}' \right) \nabla_S \cdot {\bf f}_p \left( {\bf r} \right) dS \right] dV' \\
 +  \left. \frac{1}{\varepsilon_i \left\{ 2 \omega \right\} } \iint_{T_p^r} \eta_\delta \left( { \bf r}' \right)  \left[ \iint_{ T_p^r} g_i
\left( {\bf r } -{\bf r}' \right) \nabla_S \cdot {\bf f}_p \left( {\bf r} \right) dS \right] dS' \right\}
\notag
\end{multline}
Using eqs. \ref{eq:BulkMagneticCurrent_SHG} and \ref{eq:H_Delta1} we have:
\begin{multline}
   \langle {\bf f}_p  , {\bf n} \times {\boldsymbol\pi}_{\delta'}^{\left( e \right)} \rangle  = \\
 - \sum_{r=\pm} \iiint_{V_i} {\bf J}_\delta \left( {\bf r}'
\right) \cdot \iint_{ T_p^r} \left( {\bf n} \times {\bf f}_p \right) \times \nabla g \left( {\bf r} - {\bf r}' \right) dS dV'
\notag
\end{multline}
\begin{multline}
   \langle {\bf f}_p  ,  {\boldsymbol\pi}_{\delta'}^{\left( e \right)} \rangle = \\
 - \sum_{r=\pm}   \iiint_{V_i} {\bf J}_\delta \left( {\bf r}' \right) \cdot
\iint_{ T_p^r} {\bf f}_p  \times \nabla g \left( {\bf r} - {\bf r}' \right) dS dV'
\notag
\end{multline}

\bibliographystyle{ieeetr}
\bibliography{references}

\end{document}